\documentclass[preprint,superscriptaddress,aps]{revtex4}
\usepackage{amsfonts}
\usepackage{amssymb}
\usepackage{amsmath}
\usepackage{graphics}
\usepackage{graphicx}
\usepackage{color}
\usepackage[T1]{fontenc}
\usepackage{hyperref}
\newcommand{\bea}{\begin{eqnarray}}
\newcommand{\eea}{\end{eqnarray}}

\begin{document}
\title{On the effective superpotential in the generic higher-derivative superfield supersymmetric three-dimensional gauge theory}

\author{F. S. Gama}
\email{fgama@fisica.ufpb.br}
\affiliation{Departamento de F\'{\i}sica, Universidade Federal da Para\'{\i}ba\\
 Caixa Postal 5008, 58051-970, Jo\~ao Pessoa, Para\'{\i}ba, Brazil}

\author{J. R. Nascimento}
\email{jroberto@fisica.ufpb.br}
\affiliation{Departamento de F\'{\i}sica, Universidade Federal da Para\'{\i}ba\\
 Caixa Postal 5008, 58051-970, Jo\~ao Pessoa, Para\'{\i}ba, Brazil}

\author{A. Yu. Petrov}
\email{petrov@fisica.ufpb.br}
\affiliation{Departamento de F\'{\i}sica, Universidade Federal da Para\'{\i}ba\\
 Caixa Postal 5008, 58051-970, Jo\~ao Pessoa, Para\'{\i}ba, Brazil}

\begin{abstract}
We formulate a generic three-dimensional superfield higher-derivative gauge theory coupled to the matter, which, in certain cases reduces to the three-dimensional scalar super-QED, or supersymmetric Maxwell-Chern-Simons or Chern-Simons theories with matter. For this theory, we explicitly calculate the one-loop effective potential.
\end{abstract}

\maketitle

\section{Introduction}

The effective potential is known as a central object of the quantum field theory efficiently describing its low-energy effective dynamics \cite{CW}. In the supersymmetric theories, the most adequate method for its calculation is based on the superfield formalism allowing to maintain a manifest explicit supersymmetry at all steps of the calculation. The superfield methodology for calculating the effective potential originally has been elaborated in \cite{WZ} for the four-dimensional space-time and, further, successfully applied to the Wess-Zumino model in \cite{WZ1}, to more general model involving chiral superfields only in \cite{GC}, to the supergauge theories in \cite{SYM}, and to the higher-derivative superfield theories in \cite{HD,HD1}.

However, preliminary discussions of the superfield approach to study of the three-dimensional supersymmetric field theories themselves, especially the Chern-Simons theory  \cite{CS}, and of the effective potential in three-dimensional superfield theories \cite{Burgess} began to be carried out already in 80s, in a more or less systematic way the superfield methodology to studying the effective potential has been formulated only recently \cite{ourEP,ourEP1}. However, namely recently the interest to the three-dimensional field theories has extremely grown, especially, due to the study of the $N=6$ and $N=8$ Chern-Simons theories which display finiteness and conformal invariance \cite{Mald}. Other important studies of the extended supersymmetric three-dimensional theories are presented in \cite{BS}, where, in particular, explicitly $N=2$ and $N=3$ superfield descriptions of these theories are given. Different issues related to the superfield Chern-Simons theories have been considered also in \cite{Lehum,Gall}.

However, up to now, no studies of the higher-derivative three-dimensional superfield theories were carried out whereas such a study could be certainly interesting (for example, it is natural to study the famous problem of ghosts in higher-derivative theories \cite{Smilga}, especially, taking into account that in the three-dimensional superspace the convergence is better and the formulation is simpler). The only consideration of the higher-derivative supersymmetric theories in three dimensions has been carried out within the component approach in \cite{GallWZ} where a one-loop effective potential for a model involving only scalar fields and their superpartners has been calculated.
In this work we suggest to fill this lack. We formulate the generic three-dimensional superfield higher-derivative gauge theory coupled to the matter. For this theory, we present the generic methodology for calculating the effective potential, and calculate it in an explicit way.
Throughout the paper, we follow the notations and conventions adopted in  \cite{SGRS}.  Our calculations will be carried out in Euclidean space.

\section{Higher-derivative supersymmetric gauge theory}

We start with the following three-dimensional free generic Abelian gauge theory
\bea
S=\frac{1}{2e^2}\int d^5z A^{\beta} \hat{R} D^{\gamma}D_{\beta}A_{\gamma}.
\eea
Here, $\hat{R}$ is a some scalar operator commuting with $D^{\gamma}D_{\beta}$, hence, it is a function of $D^2$, space-time derivatives and some constants. This theory is evidently invariant under usual gauge transformations $\delta A_{\alpha}=D_{\alpha}K$, with $K$ is an arbitrary scalar superfield parameter. It is clear that if (up to the multiplicative constants) $\hat{R}=1$, we have a Chern-Simons theory, if $\hat{R}=D^2$, we have a three-dimensional QED, if $\hat{R}=D^2+m$, we have a Maxwell-Chern-Simons theory. If $\hat{R}$ involves higher degrees of $D^2$, in particular, the d'Alembertian operator $\Box$ and its functions, we have the higher-derivative supersymmetric gauge theory. Earlier, the one-loop effective potential for this theory has been calculated only for the case  supersymmetric Chern-Simons theory, $\hat{R}=1$ \cite{WHD} and for the supersymmetric scalar QED, $\hat{R}=D^2$ it has been discussed in \cite{mynotes}.
We note that while the non-Abelian extension of this theory would be rather sophisticated involving the vertices of self-coupling of the gauge superfield, the one-loop effective potential will be the same as in the Abelian case, up to the constant depending on the algebraic factor, since, at the one-loop level, only the vertices involving the external scalar legs give non-trivial contributions to the effective potential.

We add to this action the following gauge-fixing term
\bea
S_{GF}=\frac{1}{2e^2\alpha}\int d^5z A^{\beta} \hat{R} D_{\beta}D^{\gamma}A_{\gamma},
\eea
which is a natural higher-derivative generalization of the usual gauge-fixing term. We note that this theory is Abelian, therefore, the ghosts completely decouple.

Now, let us couple the gauge superfield to the scalar matter. It is clear that the gauge covariant derivative is $D_{\alpha}-iA_{\alpha}$, so, the object $(D_{\alpha}-iA_{\alpha})\Phi$ (and, similarly, $(D_{\alpha}+iA_{\alpha})\bar{\Phi}$) is transformed covariantly, i.e. if one transforms $\Phi\to e^{iK}\Phi$, and takes the gauge transformation of the field $A_{\alpha}\to A_{\alpha}+D_{\alpha}K$, one will arrive at $(D_{\alpha}-iA_{\alpha})\Phi\to e^{iK}(D_{\alpha}-iA_{\alpha})\Phi$. Hence, in principle, if introduce $\nabla_{\alpha}\equiv D_{\alpha}-iA_{\alpha}$, one can introduce higher-derivative kinetic term
\bea
S^K_{\Phi}=-\frac{1}{2}\int d^5z \nabla^{\alpha}\nabla^{\beta}\ldots \nabla^{\gamma}\Phi (\nabla_{\alpha}\nabla_{\beta}\ldots \nabla_{\gamma}\Phi)^*.
\eea
We can also introduce the mass for the scalar field and the self-coupling for the scalar field, of the form $\frac{\lambda}{2}\int d^5z(\Phi\bar{\Phi})^n$. More generally, we will consider an arbitrary potential $V(\bar\Phi,\Phi)$. So, the complete action of the theory would look like
\bea
S_t&=&\int d^5z\Big[\frac{1}{2}\Big( A^{\beta}\frac{1}{e^2}\hat{R}(D^{\gamma}D_{\beta}+\frac{1}{\alpha}D_{\beta}D^{\gamma})A_{\gamma}-\nabla^{\alpha}\nabla^{\beta}\ldots \nabla^{\gamma}\Phi (\nabla_{\alpha}\nabla_{\beta}\ldots \nabla_{\gamma}\Phi)^*\Big)\nonumber\\&+&
V(\bar\Phi,\Phi)
\Big].
\eea
However, for the first attempt we suggest that the higher derivatives are present only in the gauge sector as it occurs in \cite{HD1}. Therefore, the equation above reduces to
\bea
S_t=\int d^5z\Big[\frac{1}{2}\Big( A^{\beta}\frac{1}{e^2}\hat{R}(D^{\gamma}D_{\beta}+\frac{1}{\alpha}D_{\beta}D^{\gamma})A_{\gamma}-\nabla^{\alpha}\Phi (\nabla_{\alpha}\Phi)^*\Big)+V(\bar\Phi,\Phi)
\Big].
\eea
The standard method of calculating the effective action is based on
the methodology of the loop expansion~\cite{BO}.
To do it, we make a shift $\Phi\to\Phi+ \, \phi$ in the superfield $\Phi$
(together with the analogous shift for the $\bar{\Phi}$),
where now $\Phi$ is a background (super)field, and $\phi$ is a quantum
one. We suppose the gauge field $A_{\alpha}$ is taken
to be purely quantum one. In order to calculate the effective action in the one-loop level, we have to keep only the quadratic terms in the quantum fluctuations $\phi$, $\bar\phi$ and $A_\alpha$. By using this prescription, we get
\bea
\label{quadraticaction1}
S_2[\Phi,\bar\Phi;\phi,\bar\phi,A_\alpha]&=&\frac{1}{2}\int d^5z\Big[A^{\beta} \frac{1}{e^2}\hat{R}(D^{\gamma}D_{\beta}+\frac{1}{\alpha}D_{\beta}D^{\gamma})A_{\gamma}+2\bar\phi D^2\phi+2V_{\bar\Phi\Phi}\bar\phi\phi\nonumber\\
&+&i\Phi A^\alpha D_\alpha\bar\phi-i\bar\Phi A^\alpha D_\alpha\phi+V_{\Phi\Phi}\phi^2+V_{\bar\Phi\bar\Phi}\bar\phi^2-\bar\Phi\Phi A^\alpha A_\alpha\Big] \ ,
\eea
where the irrelevant terms were omitted, including those involving covariant derivatives of the background scalar superfields. Moreover, we use a shorthand notation $V_{\bar\Phi\Phi}=\frac{\partial^2V(\bar\Phi,\Phi)}{\partial\bar\Phi\partial\Phi}$, $V_{\Phi\Phi}=\frac{\partial^2V(\bar\Phi,\Phi)}{\partial\Phi^2}$, $V_{\bar\Phi\bar\Phi}=\frac{\partial^2V(\bar\Phi,\Phi)}{\partial\bar\Phi^2}$.

From (\ref{quadraticaction1}), it follows that the propagators are given by
\bea
\label{propagators1}
\langle A_\gamma(1)A^\alpha(2)\rangle=\frac{e^2}{4k^2\hat R_1}(D_1^\alpha D_{1\gamma}+\alpha D_{1\gamma} D_1^\alpha)\delta_{12} &,& \langle \bar\phi(1)\phi(2)\rangle=\frac{D^2_1}{k^2}\delta_{12} \ ,
\eea
where $\delta_{12}\equiv\delta^2(\theta_1-\theta_2)$ is an usual Grassmannian delta function.

Now, let us study the K\"{a}hlerian potential. At one-loop order, the basic supergraphs contributing to the effective action in the theory under consideration are of three types: first, those with internal lines composed of only scalar propagators, second, those ones composed of only gauge propagators, third, those ones involving alternating gauge and matter propagators. However, as it was argued in \cite{WHD}, if we consider the Landau gauge ($\alpha=0$), then the last case do not need to be considered in our calculations, since the gauge superfield propagator $<A^{\alpha}A^{\beta}>$ in this gauge is proportional to $D^{\beta}D_{\alpha}$, while the vertex to which this propagator is associated looks like $(\Phi A^\alpha D_\alpha\bar\phi-\bar\Phi A^\alpha D_\alpha\phi)$, so, after integration by parts, the $D_{\alpha}$ acts on the propagator $<A^{\alpha}A^{\beta}>$ annihilating it due to the identity $D_\alpha D^\beta D^\alpha=0$. From now on, all the calculations presented in this work will be performed in the Landau gauge, for simplicity.

Since that the vertices $(\Phi A^\alpha D_\alpha\bar\phi-\bar\Phi A^\alpha D_\alpha\phi)$ are irrelevant in the Landau gauge, we can discard them and rewrite the functional (\ref{quadraticaction1}) as
\bea
\label{quadraticaction2}
S_2[\Phi,\bar\Phi;\phi,\bar\phi,A_\alpha]&=&\frac{1}{2}\int d^5z\Big[A^{\beta} \frac{1}{e^2}\hat{R}(D^{\gamma}D_{\beta}+\frac{1}{\alpha}D_{\beta}D^{\gamma})A_{\gamma}-\bar\Phi\Phi A^\alpha A_\alpha+\phi^i{P_i}^jD^2\phi_j\nonumber\\
&+&\phi^i{M_i}^j\phi_j\Big] \ .
\eea
where
\begin{equation}
\phi_i=\left(\begin{array}{c}
\phi\\
\bar\phi
\end{array}\right)
 \ , \ \phi^i=\left(\begin{array}{cc}
\phi \ & \ \bar\phi
\end{array}\right)
\ , \ {P_i}^j=\left(\begin{array}{cc}
0 & 1\\
1 & 0
\end{array}\right)
\ , \ {M_i}^j=\left(\begin{array}{cc}
V_{\Phi\Phi} & V_{\Phi\bar\Phi}\\
V_{\Phi\bar\Phi} & V_{\bar\Phi\bar\Phi}
\end{array}\right) \ .
\end{equation}
Therefore, the new propagators are ($\alpha=0$)
\bea
\label{propagators2}
\langle A_\gamma(1)A^\alpha(2)\rangle=\frac{e^2}{4k^2\hat R_1}D_1^\alpha D_{1\gamma}\delta_{12} &,& \langle \phi_i(1)\phi^j(2)\rangle=\frac{{P_i}^jD^2_1}{k^2}\delta_{12} \ .
\eea
These propagators will be used for the one-loop calculations.

\section{One-loop calculations}

Let us start the calculations of the one-loop supergraphs { contributing to the purely scalar sector, that is, those ones} involving the scalar superfield propagators (\ref{propagators2}) connecting the vertices $\phi^i{M_i}^j\phi_j$. Such supergraphs exhibit structures given at Fig. 1.

\begin{figure}[!h]
\begin{center}
\includegraphics[angle=0,scale=0.40]{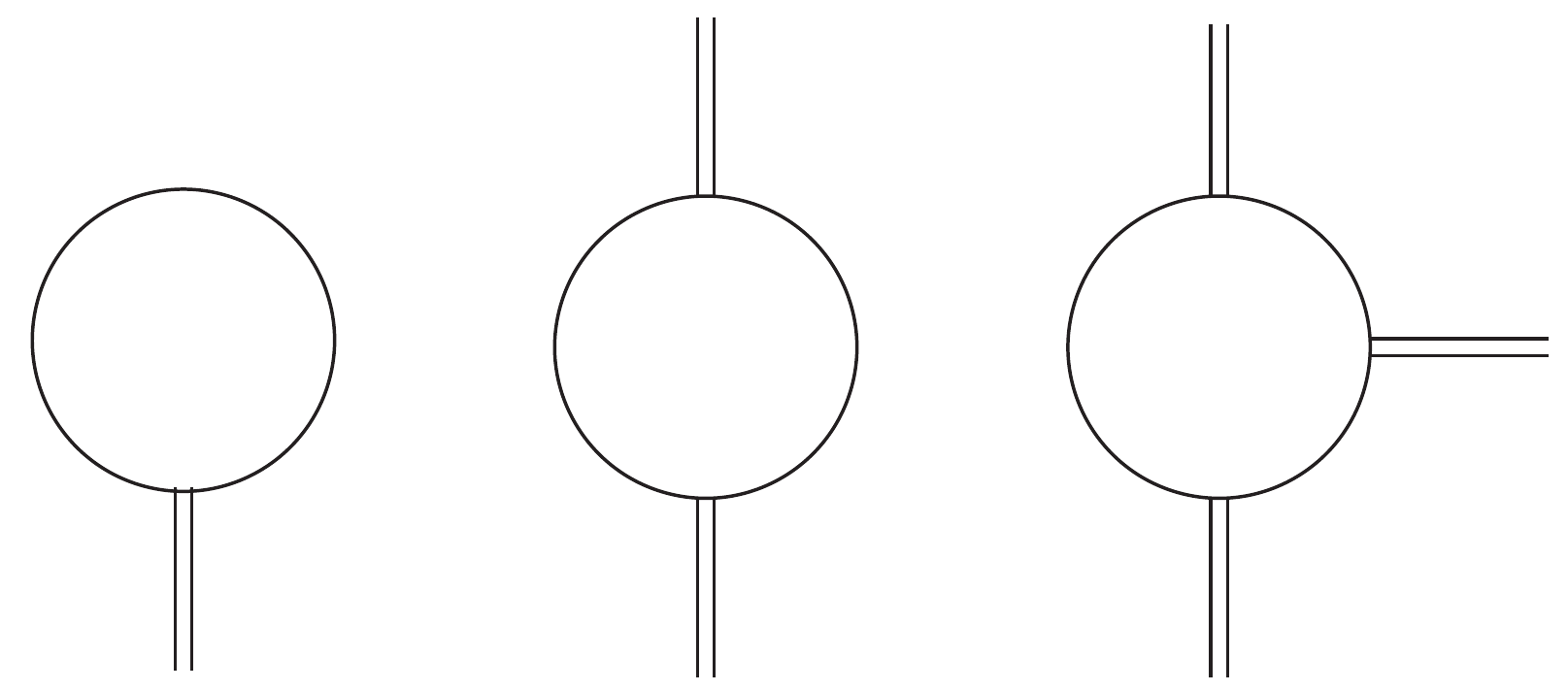}
\end{center}
\caption{One-loop supergraphs in a purely scalar sector.}
\end{figure}

We can compute all the contributions by noting that each supergraph above is formed by $n$ "subgraphs" like these ones given by Fig. 2.

\begin{figure}[!h]
\begin{center}
\includegraphics[angle=0,scale=0.70]{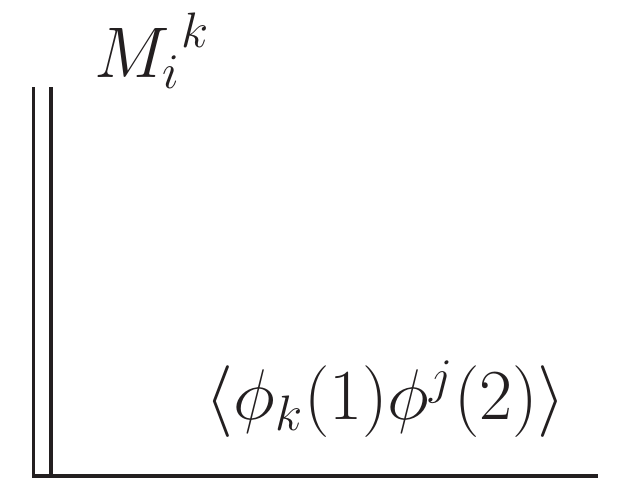}
\end{center}
\caption{A typical vertex in one-loop supergraphs in matter sector.}
\end{figure}

Hence, the contribution of this subgraph is given by
\bea
{(Q_{12})_i}^j&=&{(M_1)_i}^k{P_k}^j\frac{D^2_1}{k^2}\delta_{12}={(\widetilde M_1)_i}^j\frac{D^2_1}{k^2}\delta_{12} \ ,\\
\widetilde M&=&\left(\begin{array}{cc}
V_{\Phi\bar\Phi} & V_{\Phi\Phi}\\
V_{\bar\Phi\bar\Phi} & V_{\Phi\bar\Phi}
\end{array}\right) \ .
\eea
It follows from the result above that the contribution of a supergraph formed by $n$ subgraphs is given by
\bea
I_n&=&\int d^3x\frac{1}{2n}\int d^2\theta_1d^2\theta_2\ldots d^2\theta_{n}\int \frac{d^3k}{(2\pi)^3}\textrm{Tr}\{{(Q_{12})_i}^j{(Q_{23})_j}^k\ldots {(Q_{n-1,n})_l}^m{(Q_{n,1})_m}^p\} \nonumber\\
&=&\int d^3x\frac{1}{2n}\int d^2\theta_1d^2\theta_2\ldots d^2\theta_{n}\int \frac{d^3k}{(2\pi)^3}\textrm{Tr}\{\Big[{(\widetilde M_1)_i}^j\frac{D^2_1}{k^2}\delta_{12}\Big]\Big[{(\widetilde M_2)_j}^k\frac{D^2_2}{k^2}\delta_{23}\Big]\ldots\nonumber\\
&\times&\Big[{(\widetilde M_n)_m}^p\frac{D^2_n}{k^2}\delta_{n,1}\Big]\} \ ,
\eea
where $\textrm{Tr}$ denotes the trace over the matrix indices and $2n$ is a symmetry factor. Such a factor takes into account the Taylor series expansion coefficients of the effective action, the usual symmetry factor of each supergraph, and the number of topologically distinct supergraphs \cite{Hat}. The external momenta have to be taken to be zero in the calculation of the effective potential.

We can integrate by parts the expression $I_n$ to get
\bea
\label{supergraphs}
I_n=\int d^5z\int \frac{d^3k}{(2\pi)^3}\frac{1}{2n}\textrm{Tr}[\widetilde M^n]\Big(\frac{D^2}{k^2}\Big)^{n}\delta_{\theta\theta^{\prime}}|_{\theta=\theta^{\prime}}.
\eea
The effective action is given by the sum of all supergraphs $I_n$,
\bea
\Gamma^{(1)}_1=\sum_{n=1}^{\infty}I_n=\int d^5z\int \frac{d^3k}{(2\pi)^3}\sum_{n=1}^{\infty}\frac{1}{2n}\textrm{Tr}[\widetilde M^n]\Big(\frac{D^2}{k^2}\Big)^{n}\delta_{\theta\theta^{\prime}}|_{\theta=\theta^{\prime}} \ .
\eea
It is not difficult to prove that $(D^2)^m\delta_{\theta\theta^{\prime}}|_{\theta=\theta^{\prime}}=0$, for $m=2l$; $(D^2)^m\delta_{\theta\theta^{\prime}}|_{\theta=\theta^{\prime}}=(\sqrt{-k^2})^{m-1}$, for $m=2l+1$; where $l$ is an integer non-negative. It follows that
\bea
\label{sum}
\Gamma^{(1)}_1&=&\int d^5z\int \frac{d^3k}{(2\pi)^3}\sum_{l=0}^{\infty}\frac{(-1)^{l}}{2(2l+1)}\textrm{Tr}[\widetilde M^{2l+1}]\frac{1}{(k^2)^{l+1}}\nonumber\\
&=&\int d^5z\int \frac{d^3k}{(2\pi)^3}\sum_{l=0}^{\infty}\frac{(-1)^{l}}{2(2l+1)}[\lambda_1^{2l+1}+\lambda_2^{2l+1}]\frac{1}{(k^2)^{l+1}} \ ,
\eea
where the $\lambda$'s are the eigenvalues of the matrix $\widetilde M$, namely $\lambda_{1,2}=V_{\bar\Phi\Phi}\pm(V_{\Phi\Phi}V_{\bar\Phi\bar\Phi})^{1/2}$. Hence, substituting these eigenvalues in (\ref{sum}) and summing over all $l$ we get
\bea
\Gamma^{(1)}_1&=&\frac{1}{2}\int d^5z\int \frac{d^3k}{(2\pi)^3}\frac{1}{|k|}\Big[\arctan\Big(\frac{V_{\bar\Phi\Phi}+(V_{\Phi\Phi}V_{\bar\Phi\bar\Phi})^{1/2}}{|k|}\Big)+
\nonumber\\&+&
\arctan\Big(\frac{V_{\bar\Phi\Phi}-(V_{\Phi\Phi}V_{\bar\Phi\bar\Phi})^{1/2}}{|k|}\Big)\Big] \ .
\eea
Finally, we can compute these integrals to get
\bea
\label{part1}
\Gamma^{(1)}_1=-\frac{1}{8\pi}\int d^5z(V_{\bar\Phi\Phi}^2+V_{\Phi\Phi}V_{\bar\Phi\bar\Phi}) \ .
\eea
We conclude here that this contribution to the one-loop effective action does not display any divergences, independently of the form of the potential $V(\bar\Phi,\Phi)$. We note that, unlike \cite{WHD}, here we used only the supergraph summation, instead of the functional trace calculations.

Let us move on to the calculation of the one-loop supergraphs involving the gauge superfield propagator connecting the vertices $-\bar\Phi\Phi A^\alpha A_\alpha$. Such supergraphs exhibit a structure given by Fig. 3.

\begin{figure}[!h]
\begin{center}
\includegraphics[angle=0,scale=0.40]{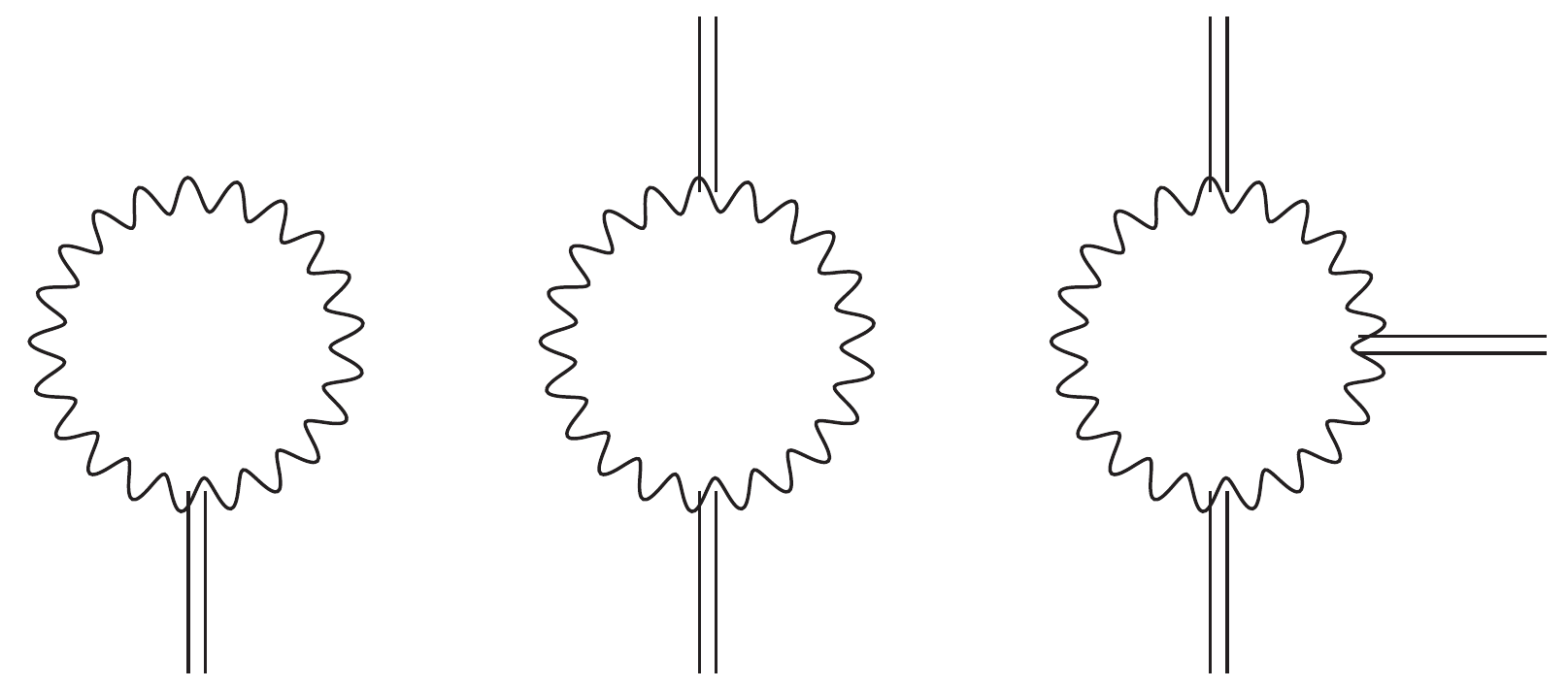}
\end{center}
\caption{One-loop supergraphs in a gauge sector.}
\end{figure}
\newpage
As before, we can compute all the contributions by noting that each supergraph above is formed by $n$ subgraphs like these ones depicted at Fig. 4.

\begin{figure}[!h]
\begin{center}
\includegraphics[angle=0,scale=0.70]{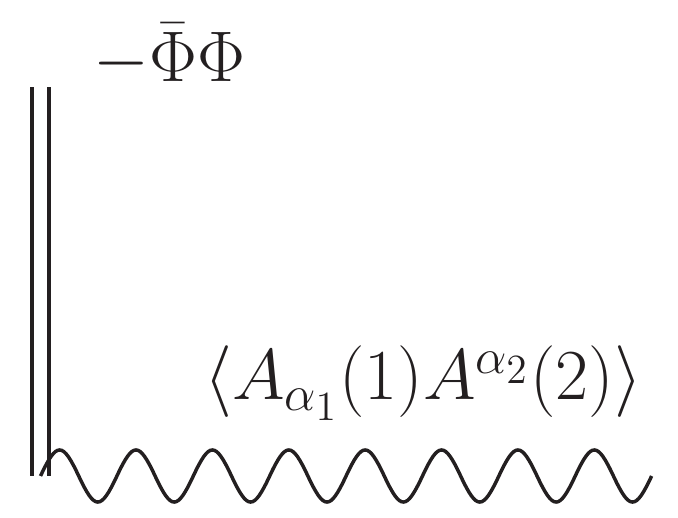}
\end{center}
\caption{A typical vertex in one-loop supergraphs in matter sector.}
\end{figure}

This subgraph provides the contribution
\bea
{(P_{12})_{\alpha_1}}^{\alpha_2}=-\frac{e^2(\bar\Phi\Phi)_1}{4k^2}\frac{1}{\hat R_1}D^{\alpha_2}_1D_{1,\alpha_1}\delta_{12} \ .
\eea
It follows from the result above that the contribution of a supergraph formed by $n$ subgraphs is given by
\bea
J_n&=&(2\pi)^3\delta^3(0)\frac{1}{2n}\int d^2\theta_1d^2\theta_2\ldots d^2\theta_n\int \frac{d^3k}{(2\pi)^3}{(P_{12})_{\alpha_1}}^{\alpha_2}{(P_{23})_{\alpha_2}}^{\alpha_3}\ldots{(P_{n-1,n})_{\alpha_{n-1}}}^{\alpha_n}
{(P_{n,1})_{\alpha_n}}^{\alpha_1}\nonumber\\
&=&(2\pi)^3\delta^3(0)\frac{1}{2n}\int d^2\theta_1d^2\theta_2\ldots d^2\theta_n\int \frac{d^3k}{(2\pi)^3}\Big[-\frac{e^2(\bar\Phi\Phi)_1}{4k^2}\frac{1}{\hat R_1}D^{\alpha_2}_1D_{1,\alpha_1}\delta_{12}\Big]\nonumber\\
&\times&\Big[-\frac{e^2(\bar\Phi\Phi)_2}{4k^2}\frac{1}{\hat R_2}D^{\alpha_3}_2D_{2,\alpha_2}\delta_{23}\Big]\ldots
\Big[-\frac{e^2(\bar\Phi\Phi)_n}{4k^2}\frac{1}{\hat R_n}D^{\alpha_1}_n D_{n,\alpha_n}\delta_{n,1}\Big] \ .
\eea
After successive integration by parts and summing all supergraphs $J_n$, we get the effective action
\bea
\label{stop}
\Gamma^{(1)}_2&=&\int d^5z\int\frac{d^3k}{(2\pi)^3}\sum_{n=1}^{\infty}\frac{1}{2n}\Big(-\frac{e^2\bar\Phi\Phi}{4k^2}\Big)^n\frac{1}{\hat R^n}
D^{\alpha_2}D_{\alpha_1}D^{\alpha_3}D_{\alpha_2}\ldots D^{\alpha_n}D_{\alpha_{n-1}}\times\nonumber\\&\times&
D^{\alpha_1}D_{\alpha_n}\delta_{\theta\theta^{\prime}}|_{\theta=\theta^{\prime}} \ .
\eea
At this stage of the calculation, we have to specify the operator $\hat R$  in order to proceed with the calculation of $\Gamma^{(1)}_2$. The most general choice is $\hat R=f(\Box)+g(\Box)D^2$ (remind that this operator is a scalar). 
This expression is rather generic. The result of the complete evaluation of the D-algebra essentially depends on the explicit form of the operator $\hat{R}$. So, let us consider two characteristic examples where the final result is expressed in closed form and in terms of elementary functions.

The first example is $f=0$ and $g\ne0$, so we have
\bea
\label{firstchoice}
\hat R=g(\Box)D^2\Rightarrow\frac{1}{\hat R^{n}}=\bigg(\frac{-1}{g(k^2)k^2}\bigg)^{n}(D^2)^n \ .
\eea
It follows from the covariant derivatives algebra that $(D^2)^nD^{\alpha_2}D_{\alpha_1}D^{\alpha_3}D_{\alpha_2}\ldots D^{\alpha_1}D_{\alpha_n}\delta_{\theta\theta^{\prime}}|_{\theta=\theta^{\prime}}=0$ for all $n$. Therefore, from the Eqs. (\ref{stop}) and (\ref{firstchoice}), we have
\bea
\label{part2.1}
\Gamma^{(1)}_2=0 \ .
\eea
In conclusion, the complete one-loop K\"{a}hlerian effective potential { is completely given by the expression} (\ref{part1}):
\bea
K^{(1)}(\Phi,\bar\Phi)=-\frac{1}{8\pi}(V_{\bar\Phi\Phi}^2+V_{\Phi\Phi}V_{\bar\Phi\bar\Phi}) \ , \ &\textrm{for}& \ f(\Box)=0 \ {\rm and} \ g(\Box)\ne0 \ .
\eea
This result is consistent with the claim made in \cite{mynotes} that, in the case of the absence of the self-coupling of the scalar field, the one-loop K\"{a}hlerian effective potential for the three-dimensional QED (that is, $g=1$) identically vanishes.
We showed that the same situation occurs for the all class of theories in which $g\ne1$, but $f=0$.

Our second example is $f=\xi(-\Box)^m$, and $g=0$, where $\xi$ is a parameter with a non-trivial mass dimension $[\xi]=[M]^{-2m}$, $\xi>0$, and $m$ is a non-negative integer. Consequently, we have trivially
\bea
\label{secondchoice}
\hat R=\xi(-\Box)^m\Rightarrow\frac{1}{\hat R^{n}}=\bigg(\frac{1}{\xi(k^2)^m}\bigg)^{n} \ .
\eea
It can be shown that $D^{\alpha_2}D_{\alpha_1}D^{\alpha_3}D_{\alpha_2}\ldots D^{\alpha_1}D_{\alpha_n}\delta_{\theta\theta^{\prime}}|_{\theta=\theta^{\prime}}=0$, for $n=2l$; $D^{\alpha_2}D_{\alpha_1}D^{\alpha_3}D_{\alpha_2}\ldots D^{\alpha_1}D_{\alpha_n}\delta_{\theta\theta^{\prime}}|_{\theta=\theta^{\prime}}=2^n(\sqrt{-k^2})^{n-1}$, for $n=2l+1$; where $l$ is an integer non-negative. Hence, from the Eqs. (\ref{stop}) and (\ref{secondchoice}), we get
\bea
\Gamma^{(1)}_2&=&\int d^5z\int\frac{d^3k}{(2\pi)^3}\frac{1}{\sqrt{-k^2}}\sum_{l=0}^{\infty}\frac{1}{2(2l+1)}\bigg(-\frac{e^2\bar\Phi\Phi\sqrt{-k^2}}{2\xi (k^2)^{m+1}}\bigg)^{2l+1}\nonumber\\
&=&-\frac{1}{2}\int d^5z\int\frac{d^3k}{(2\pi)^3}\frac{1}{|k|}\arctan{\bigg(\frac{e^2\bar\Phi\Phi|k|}{2\xi(k^2)^{m+1}}\bigg)} \ ,
\eea
where we have used the fact that $\sqrt{-k^2}=i|k|$ and the identity $\arctan(x)=\frac{1}{i}\textrm{arctanh}(ix)$. The integral above can be solved by induction. Then we obtain
\bea
\label{part2.2}
\Gamma^{(1)}_2=-\int d^5z\frac{1}{16\pi}\sec\bigg(\frac{\pi}{2m+1}\bigg)\bigg(\frac{e^2\bar\Phi\Phi}{2\xi}\bigg)^{\frac{2}{2m+1}} \ .
\eea
It is worth to notice that this result is finite and does not need any renormalization, { which, however, is  rather generic effect in the three-dimensional superfield theories \cite{mynotes}. We note that, moreover, if the operator $\hat{R}$ is of the first order in space-time derivatives or, as is the same, of the second order in spinor supercovariant derivatives, the theory is super-renormalizable, with the only possible divergences are the two-loop ones, and no divergences at higher order, this is just the situation of the super-QED \cite{qedfin}, and if the operator $\hat{R}$ is of second order in space-time derivatives, the corresponding theory is all-loop finite.}.

Again, the complete one-loop K\"{a}hlerian effective potential can be read off from the sum of Eqs. (\ref{part1}) and (\ref{part2.2}). As a result, we finally obtain
\bea
\label{finalresult}
K^{(1)}(\bar\Phi,\Phi)&=&-\frac{1}{16\pi}\sec\bigg(\frac{\pi}{2m+1}\bigg)\bigg(\frac{e^2\bar\Phi\Phi}{2\xi}\bigg)^{\frac{2}{2m+1}}
-\frac{1}{8\pi}(V_{\bar\Phi\Phi}^2+V_{\Phi\Phi}V_{\bar\Phi\bar\Phi}) \ ,
\eea
for $f(\Box)=\xi(-\Box)^m$ and $g(\Box)=0$.

The result (\ref{finalresult}) is highly generic. In particular, if $m=0$, $\xi=1$, and $V(\bar\Phi\Phi)=\frac{\lambda}{2}(\bar\Phi\Phi)^2$ we get
\bea
\label{without}
K^{(1)}(\bar\Phi,\Phi)&=&\frac{1}{64\pi}(e^2\bar\Phi\Phi)^{2}-\frac{5}{8\pi}\lambda^2(\bar\Phi\Phi)^2 \ .
\eea
This is just the (Euclidean) one-loop K\"{a}hlerian effective potential for the Chern-Simons theory coupled to a self-interacting massless scalar matter without higher derivatives. Our result agrees with that one obtained in \cite{WHD}.

\section{Summary}

We formulated a generic Abelian three-dimensional supergauge theory coupled to a matter. In a general case, the classical action of this theory involves higher derivatives. However, despite of this, we developed an universal procedure for calculating the one-loop effective potential for this theory, that is, actually, or a wide class of theories including supersymmetric Chern-Simons theory, supersymmetric QED, supersymmetric Maxwell-Chern-Simons theory and their non-Abelian generalizations and found that the result is rather generic for a wide class of the theories. In particular, we explicitly demonstrated that for the three-dimensional supersymmetric QED the one-loop effective potential vanishes. Also, we note that any three-dimensional higher-derivative supersymmetric gauge theory is all-loop finite.

The studies of the higher-derivatives supefield theories can have natural continuation.
For example, it is interesting to look for another manners to introduce higher-derivatives in the four-dimensional superfield theories, different from that ones presented in \cite{HD,HD1}.

\vspace{5mm}

{\bf Acknowledgments.}
This work was partially supported by Conselho Nacional de
Desenvolvimento Cient\'\i fico e Tecnol\'ogico (CNPq). A. Yu. P. has
been supported by the CNPq project No. 303438-2012/6. The work by F. S. Gama has been supported by the
CNPq process No. 141228/2011-3.

\end{document}